\documentstyle[aaspptwo]{article}
\begin{document}

\title{ RADIATION INDUCED VOID IN THE SPECTRUM OF TOL~1038-2712}

\author{R. Srianand \\
        Inter-University Centre for Astronomy and Astrophysics \\
        Post Bag 4, Ganeshkhind, Pune 411 007, India,\\
	email anand@iucaa.ernet.in}

\begin{abstract}
Detection of a large void ($\sim $7 Mpc) is reported between the
redshifts 2.16286 and 2.20748 in the Ly~$\alpha$ forest of TOL
1038-2712. This void is centered near a foreground QSO TOL 1037-2704
which is at a distance $\sim$4.4 Mpc away from the void. The estimated
probability for the void to occure by chance in front of the foreground
QSO is few times $10^{-3}$. Various implications of the void being
produced by excess ionization due to foreground QSO are discussed.
\vskip0.4in
\noindent 
{\it Subject headings:}
QSO:general-QSO:absorption lines-IGM-intergalactic medium.
\end{abstract}
\section {INTRODUCTION}
Ly~$\alpha$ lines seen in the spectra of QSOs are believed to be
produced by neutral hydrogen clouds in the intergalactic medium (IGM).
Being the most abundant objects at higher redshifts Ly~$\alpha$ clouds can
provide valuable information regarding the IGM over a wide span of
look back time.  Bajtlik et al. (1998, here after BDO) have shown that
the deficit of lines seen in the Ly~$\alpha$ forest close to the
QSOs, known as proximity effect, can be used to estimate the intensity
of the ionizing UV background radiation.

Properties of Ly~$\alpha$ clouds in addition to be useful in understanding
the IGM can be used to infer various physical properties of QSOs.
Srianand \& Khare (1996) have shown Ly$\alpha$ clouds with ${\rm z_{abs}>
z_{em}}$ can be used to get bounds on the relative velocity shifts between
broad emission lines, peculiar velocity of the QSOs and their evolution
with redshift.

In this work we study the effect of foreground QSO TOL 1037-2704 on the
Ly~$\alpha$ forest of the background QSO TOL 1038-2712. We have used
Ly~$\alpha$ line list from the sample of Dinshaw \& Impey (1996)
which is originally compiled to study the extent of large scale structure of
the  intervening metal line absorbers. We have identified a void
exactly at the position of the foreground QSO and estimated its
significance using the numerical simulations. The implications of void
on the estimates of background radiation and the QSO models are discussed
in the following sections. For all our calculations we use 
${\rm H_o =100\;km\;s^{-1}\;Mpc^{-1}}$ and ${\rm q_o = 0.5}$.

\section {VOID IN THE Ly~$\alpha$ FOREST}

The distribution of Ly~$\alpha$ clouds and QSOs in the field of TOLOLO
group of QSOs are shown in Figure 1.  One can clearly see an apparent
lack of Ly~$\alpha$ absorption lines along the line of sight to Tol
1038-2712 (between redshifts 2.1652 and 2.2075) close to the foreground
QSO Tol 1037-2704 (${\rm z_{em}= 2.195}$).  The angular seperation
between the two QSOs is 17'.9.  This corresponds to a proper separation
of 4.4${\rm h^{-1}\;Mpc}$ at z = 2.195.  The redshift difference 0.0422
at z = 2.195 gives a proper length  of  $\sim 7{\rm h^{-1}\; Mpc}$ for
the void.

We have estimated the number of Ly~$\alpha$ clouds per unit redshift,
N(z), for the Ly~$\alpha$ lines along 4 QSO sight lines, which are
3 Mpc away from the QSOs, using maximum
likelihood method. The results for limiting rest equivalent width
${0.1\AA}$ together with redshift distribution calculated based on the
high resolution sample compiled by Srianand \& Khare (1994, sk94) are
given in Table 1. Also given in the table are the total number of lines
used in the analysis, total redshift path and average redshift of the
samples. We performed numerical simulations to estimate the
significance for the occurence of a gap near the foreground QSO to
establish a connection between the void and the foreground QSO.

In our numerical simulations we assume  no evolution in the number
density of Ly~$\alpha$ clouds as this is true for small redshift
intervals. The distribution of number of Ly~$\alpha$ clouds along the
line of sight is assumed  to be Poissonian with  mean given in Table
1.  We generated a random line list for the QSO Tol 1038-2712
considering the observed redshift window. We repeated the simulations
for over 50000 times for different values of the number of Ly~$\alpha$
clouds per unit redshift (as given in Table 1). The estimated
probability of occurence of voids, with size greater than the observed
one (and number of lines along the line of sight same as the observed
number) anywhere along the line of sight and close to the foreground
QSO are given in Table 2 (P1 and P2 respectively).  The
probability of finding a gap similar or greater than the observed void
by chance is few times 10$^{-3}$. Thus it is clear the void is real and
not due to any random fluctuations in the interline spacing of
Ly~$\alpha$ clouds. In what follows we discuss various implications
of the presence of this void.
\section{IMPLICATIONS OF THE VOID}
\subsection{Background Radiation field}
The presence of foreground QSO close the centre of the void naturally
favours the void being produced by the proximity effect. If this is
true one can get a bound on the intensity of ionizing UV background
radiation, ${\rm J_\nu}$. We have assumed the QSO radiation to be
emitted isotropically and has not varied much over a long time scale.
We have calculated the continuum flux at the Lyman limit for each QSO
in the field from its V-magnitude assuming the spectral energy
distribution of the QSOs to be a powerlaw, $f_\nu\propto
\nu^{-\alpha}$,  (with spectral index $\alpha=0.636\pm0.303$ estimated
from Tytler \& Fan 1992). We have not introduced any correction to the
magnitude due to emission line flux as estimated contribution in the
redshift window 2.0 to 2.5 is very low (between 0.005 to 0.025).

We have plotted, in Figure 2, the total Lyman limit flux seen by each
Ly~$\alpha$ clouds due to all the know QSOs in the field against the
absorption redshift. We have not taken the absorption along the line of
sight into account in the calculations as the redshift path lengths are
small.  The quoted error bars are errors in flux due to 1$\sigma$
uncertainty in the mean spectral index.  Also plotted in the figure are the
background radiation values quoted by various authors in the
literature.

The Lyman limit flux at the center of the void is ${\rm log(f/4\pi)}
\;=\;{-21.60}^{+0.04}_{-0.03}$ ${\rm
ergs\;s^{-1}\;cm^{-2}\;Hz^{-1}\;Sr^{-1}}.$  Thus the implied upper
limit of the background radiation field at z$\sim$2.19 is,
\begin{equation} 
{\rm log (J_\nu)\; \le {-21.60}^{+0.04}_{-0.03}},
\end{equation}
in ${\rm ergs\;s^{-1}\;cm^{-2}\;Hz^{-1}\;Sr^{-1}}$.  This value is
consistent with the lower limit, -21.80  ${\rm
ergs\;s^{-1}\;cm^{-2}}$ ${\rm Hz^{-1}\;Sr^{-1}},$ quoted by Fernandez-Soto et
al (1995) based on the proximity effect on the Ly~$\alpha$ forest due
to foreground QSOs. It is clear from the figure 2 that the quoted upper
limit is much lower than most of the values available in the literature
based on the proximity effect.

We have performed the standard proximity effect calculations prescribed
by BDO using the available Ly~$\alpha$ data  along the 4 QSO sight
lines in the field. We have calculated the expected number of lines in
different relative velocity bins with respect to QSOs for column density
distribution index, $\beta = 1.5$,  and rest equivalent width cutoff
0.1$\AA$. In these calculations we have assumed no evolution for the
number density of Ly~$\alpha$ clouds and used the mean number densities
given in Table 1. The ratio of expected number, ${\rm N_{exp}}$,  to
observed number, ${\rm N_{obs}}$, of Ly~$\alpha$ clouds in different
relative velocity bins with respect to the QSOs are given in Figure 3
for three different values of the background intensity. The errorbars
are calculated considering the distribution of number of lines along the
line of sight to be Poissonian.

I-model underpredicts the number of Ly~$\alpha$ clouds near the QSO by
more than 2$\sigma$ when we consider the background intensity
value equal to the obtained upper limit (top pannel in Figure 3). The
predicted distribution is however consistent with the background value
estimated by BDO and Bechtold (1994)(bottom pannel in Figure 3). The
disagreement between the value of background intensity calculated using
two different method can be accounted for if either our assumption of
isotropic emission is wrong or the QSO luminosity has varied within the
light travel time between QSO and void.  In the following sections we
discuss the implications of these possiblities in detail.
\subsection {Long term fading of QSOs}
In this section we assume the emisson from the QSO to be isotropic and
try to put constraints on different QSO evolution models based on the
background radiation values estimated from the void and from general
proximity effect calculations. One can estimate lower limit on the life
time of the QSO from its distance from the void.  The separation 4.4
Mpc corresponds to a light travel  time of 2$\times10^7$ Yrs. Thus the
life time of the foreground QSO 1037-270 is
\begin{equation}
t_{\rm qso}\;\ge\;2\times10^7\; {\rm Yrs.}
\end{equation}
The observed luminosity evolution of QSOs can be interpreted as (a) the
evolution of a single long lived ($t_{\rm qso}=10^{10}$ Yrs) population
of QSOs or (b) the evolution of successive geneation of short lived
QSOs ($t_{\rm qso}=10^8$ Yrs).  Our estimated lower limit is consistent
with both the models for QSOs. Boyle et al.(1991) have shown that the
luminosity evolution of QSOs can be parameterized by
\begin{equation}
L(z)\;\propto\;(1+z)^k
\end{equation}
with best fitted value of $k\;=\;3.45\pm0.10$. Thus in the case of long
lived population the luminosity of QSOs will not change within the
light travel time between TOL 1037-270 and void. Both the models are
consistent if the background intensity is less than the upper limit
quoted based on the presence of void in the previous section. However
if the actual background value is higher than the upper limit quoted
here, as suggested by the general proximity effect estimates, one will
need a reduction in the luminosity of TOL 1037-270 by a factor $>2-16$
within 2$\times10^7$ Yrs. Dobrzycki \& Bechtold (1991) identified a
void in the spectra of Q0302-003 close to a foreground QSO Q0301-005.
Unlike the present case the void is displaced from the foreground QSO
towards lower wavelength side. The distance between the centre of the
void and  Q0301-005 is 4.25 Mpc. The corresponding light travel time is
1.34 $\times10^7$ Yrs. The estimated Lyman limit flux from the QSO at
the center of the void is $10^{-21.8}$ ${\rm
ergs\;s^{-1}\;cm^{-2}\;Hz^{-1}\;Sr^{-1}}$.  Thus for different
background value estimated in the literature Q0301-005 (like TOL
1037-270) also needs a reduction in the luminosity by a factor $>3-40$
within 2$\times10^7$ Yrs. Fadding of QSOs within such a short time
scale can not be explained by the long lived QSO models.  Thus if the
intensity of background radiation is high and if QSOs radiate
isotropically, observation of voids due to foreground QSOs favour models in
which the luminosity evolution is realized as the superposition of
activities in successive generation of short lived QSOs (Hachnelt \& Rees 1993).

It is known that long lived interpretation has problems with
observations. It predicts larger remnant black holes ($\sim 10^9{\rm
M_\odot}$) and smaller Eddington ratios ($\sim0.001$) in low redshift
AGNs than are corrently inferred from emission line and continuum
studies (Padovani 1989). Our analysis gives an independent observational proof
against the long lived QSO interpretation.
\subsection{Anisotropic Emission in QSOs}

Crotts (1989) has suggested an anisotropic emission from QSOs to
explain the nondetection of proximity effect due to foreground QSOs.
If TOL 1037-2704 is not an isotropic emitter and not varied, then we
require an excess collimated beam (with radiation roughly an order of
magnitude more compared to that along our line of sight) within a cone
of angle 76$^{\rm o}$ perpendicular to the line of sight.

Dorbrzycki \& Bechtold (1991), in order to explain the displaced void,
proposed an opening angle of 140$^{\rm o}$ assuming the radiation towards
void is same as that received along our line of sight. If their
assumption is true one should see a void in the spectrum of Q0301-005
close to the emission redshift along our line of sight. We are not
finding any such deficit in the intermediate resolution spectra of
Q0301-005 observed by Steidel (1990) and simple assumption of
Dorbrzycki \& Bechtold (1991) may not be correct.  The distribution of
lines near this QSO is consistent with high values of background
radiation.  The void can be explained if the excess collimated beem is
within a cone angle of $<70^{\rm o}$. Thus the foreground void in both
the cases favour a narrow collimated beam inexcess to a isotropic
emission than a wide angle cone of emission caused by shadowing of a
portion of the emission.  However more observations are
needed to get a clear picture about the beaming models.

\section{CONCLUSIONS}

We have reported detection of a void ($\sim 7 {\rm Mpc}$) in the
spectrum of QSO TOL 1038-2712 close to the foreground QSO TOL
1037-2704. Using numerical simulations we have shown the chance
probability of occurence of void close to the foreground QSO is
$\sim10^{-3}$.

We have estimated the upper limit on intensity of background radiation
field to be $ {-21.6}^{+0.04}_{-0.03}$ ${\rm ergs\; s^{-1}}$ ${\rm
cm^{-2}\; Hz^{-1}\; Sr^{-1}}$ assuming the void is produced by the
proximity effect of the foreground QSO. Based on the I-model and
Ly~$\alpha$ lines along four QSO sightlines we have shown the actual
background intensity is higher than the estimated upper limit. We
discuss two possiblities inorder to account for this difference.

If QSOs emit isotropically the presence of void will give another
observational proof against long lived population models of QSOs. If
QSO does not vary within few times $10^7$ Yrs, our result suggest a
narrow collimated cone emission toward the void (with flux a order or
magnitude higher) in addition to the isotropic emission to be a
possible source of the void.

Most of the available observations of pairs of QSO spectra till date
(Crotts 1990; Fernendez-Soto et al. 1996) fail to detect voids near
foreground QSOs. Even in the only two cases where voids are detected
one is found at the redshift of the foreground QSO and the other at
redshift less than that of the foreground QSO. If beaming picture is
correct one would like to see displaced voids towards higher wavelenght
side also. Thus case with displace void towards higher wavelength side
will give a proof for the beaming arguments.
%\eject
\vskip 0.4in
%%%%%%%%%%%%%%%%%%%%
%    References    %
%%%%%%%%%%%%%%%%%%%%

%%%%%%%%%%%%%%%%%%%%%
%       TABLES      %
%%%%%%%%%%%%%%%%%%%%%
%\newpage
\begin{table}[h]
\centerline{ Table 1: Results of Maximum Likelihood Analysis}
\centerline{\begin{tabular}{ccccc}
\tableline
\tableline
\multicolumn {1}{c}{Sample}&\multicolumn{1}{c}{Number}&
\multicolumn {1}{c}{z-path}&${\rm <z>}$&\multicolumn{1}{c}{N(z)}\\
\tableline
Whole    & 79&0.60&2.30&132.22$\pm$11.50\\
SK94     &374&2.24&2.73&152.31$\pm$12.31\\
\tableline
\end{tabular}}
\end{table}
%\newpage
\begin{table}[h]
\centerline{Table 2: Results of Numerical Simulations}
\centerline{\begin{tabular}{ccc}
\tableline
\tableline
N(z)&P1($10^{-3}$)&P2($10^{-3}$)\\
\tableline
152.31&2.7&1.1\\
141.99&3.3&1.5\\
132.22&5.0&2.0\\
\tableline
\end{tabular}}
\end{table}
\newpage
%%%%%%%%%%%%%%%%%%%%%%%%%%%%%figures%%%%%%%%%%%%%%%%%%%%%%%
\centerline{\bf Figure Caption}
\vskip 0.1in
\noindent{ Fig 1. Distribution of QSOs and Ly~$\alpha$ clouds in the field
of Tololo group of QSOs. Ly~$\alpha$ clouds are represented by open squares
and the void under consideration is marked as "V".}

\vskip 0.1in
\noindent{Fig 2. Total UV ionizing flux, due to all know QSOs in the field,
received by Ly~$\alpha$ absorbing clouds along four lines of sights. Error bars
are due to error in average spectral index. Horizondal lines are available
background radiation estimates in the literature.}

\vskip 0.1in
\noindent{Fig 3. Ratio of expected number, ${\rm N_{exp}}$, to the observed number, ${\rm N_{exp}}$, of Ly~$\alpha$ 
clouds in different relative velocity bins with respect to QSOs (calculated
for N(z)=152.31). The values of ${\rm J_\nu}$ are given in ${\rm ergs\; s^{-1}}$ ${\rm cm^{-2}\; Hz^{-1}\; Sr^{-1}}$.

\end{document}